# POLICING AND GATEKEEPING IN STEM
## SAFETY, SECURITY & SANCTITY

Apriel K Hodari,[1] Shayna B Krammes[1]
Chanda Prescod-Weinstein,[2] Brian D Nord,[3] Jessica N Esquivel,[3] Kétévi A Assamagan[4]

[1]Eureka Scientific Inc, Oakland, CA 94602
[2]University of New Hampshire, Durham, NH 03824
[3]Fermi National Accelerator Laboratory, Batavia, IL 60510
[4]Brookhaven National Laboratory, Upton, NY 11973

EXECUTIVE SUMMARY

The purpose of this white paper is to lay out the impacts of policing and gatekeeping in STEM, illustrated with lived experiences of scientists of color who are achieving despite the daunting challenges they face.

*Policing*

People around the world were impacted by the extrajudicial murders of Ahmaud Arbery, Breonna Taylor, and George Floyd. The effect on black people, including black scientists, was profound. In this paper, we described direct experiences black scientists have had with policing, as well as the trauma black scientists experience each time a murder like this is reported. This suffering is compounded when colleagues and peers seem oblivious and unaffected, leaving black scientists further isolated in an already unwelcoming environment.

*Gatekeeping*

In practice, gatekeeping comprises a set of behaviors, practices, and traditions, backed up by individual and organizational power to guard the boundaries of the discipline. Unfortunately, many people who bear the brunt of systemic oppression, receive multiple messages that they do NOT belong. For some, these accumulate to push them firmly outside of the boundaries, and they leave.

Even when gatekeeping fails to achieve its ultimate goal, smaller encounters exact time and emotional labor from the targets of oppression, reducing the time and energy they have available for their scientific work. Further, biases that impact how scientists efforts are judged have led to exclusions from opportunities and funding, which lead to further losses.

*Comfort and Safety*

We invite readers to wrestle with the difference between feeling unsafe and actually being unsafe. Using the experiences of real people, we describe productive enactments of this tension, and reveal the benefits of accepting this struggle as ongoing and endless.

*Take-Aways*

The paper concludes with an account of how even a well-intentioned, self-described social activist can cause harm, contrasted against someone working daily to create an inclusive environment for everyone to work and learn.

INTRODUCTION

In 2018, Lolade Siyonbola, an African Studies graduate student at Yale University, took a break from studying for finals, and laid down for a nap in a common room of her campus dorm. She was awakened at 1:40 am by fellow graduate student Sarah Braasch and told she couldn't sleep there. Ms. Braasch then called campus police. Ms. Siyonbola filmed and shared her interactions with Braasch and campus police (Gerken, 2018; Rogo, 2020). She later described her decision to film the encounters as motivated in part by her fear that she would suffer the fate of motorist Sandra Bland (Laughland, 2019). The video of Siyonbola's experience went viral on social media, and inspired intense conversations about the policing of shared institutional space. As Professor Eddie S Glaude, Jr articulated about students at his own university (2020):

> *Black students at Princeton aren't interlopers. They aren't guests on campus or the beneficiaries of charity who should be grateful to be at the school. They are an integral part of the Princeton community, … And, like the other students on campus, they should feel a sense of possession of the university.*

Braasch felt so much possession of the common space in their shared dorm that she was entitled to control who used the space and how they used it. Siyonbola not only felt that her right to be there was questioned, but that her life might be taken as a consequence.

These differential senses of belonging and possession, and the resultant agency each student enacted become key features not only of their individual graduate experiences, but they become features *of the setting* for these students. Siyonbola could not characterize her Yale experience as different from her interaction with Braasch and the campus police. She carried the knowledge that she needed to justify her presence in the very building in which she slept every night everywhere she went. That knowledge did not disappear when she crossed the threshold into classrooms, or the library, or any other building on campus. The threat Siyonbola felt was a Yale threat. And unfortunately, the administration's vague initial response compounded fear with frustration.

As scholar Robin DiAngelo articulates, in the cultural context created by *the lie*, those in power in an oppressive society perceive status-based experiences like Siyonbola's as normal (DiAngelo, 2018). In DiAngelo's words, "it is comfortable for me, as a white person, to live in a racist society." To change this dynamic, and thereby systemic oppression in physics, all of us have to get used to feeling uncomfortable, rather than just the targets of oppression. Further, we must build systems, practices, and structures to keep us uncomfortable with oppression.

*I*n this paper, we will explore the impacts of policing and gatekeeping on people on the losing side of power dynamics in physics. We begin with an examination of how systemic racism affects black scientists. Next, we explore how common behaviors, beliefs, and practices serve to keep physics largely white and male. In the last section, we distinguish between comfort and safety, and invite you to examine what the new knowledges presented in this paper commend to us to do about it.

## Black Scientists will be Safe when Black People are Safe

Dr. Charles D. Brown II is a postdoctoral researcher of ultracold atoms at University of California Berkeley, a 2020 Ford Foundation fellow, and the 2021 recipient of the Quantum Creators Prize from the University of Chicago. He is a director and professional events and communications advisor for Black in Physics, an organization "dedicated not only to celebrating Black physicists and our contributions to the scientific community, but also to reveal a more complete picture of what a physicist looks like" (Brown et al.). Black in Physics has an annual, week-long event, as well as featured articles in *Physics Today.*

Neither Dr. Brown's identity as a physicist, nor his many achievements, can negate the fact that he is a black man; he is someone who once experienced police violence because of the color of his skin. In July 2020, Dr. Charles Brown published an essay in *Physics Today* entitled, "Disentangling anti-Blackness from physics" (Brown II, 2020). In it, Brown describes the impact policing and gatekeeping had on him, as he has navigated his education and career with our discipline. Surrounded by the gaslighting belief that the purity of our science makes physics "a culture of no culture", Brown had to make the painful choice between minimizing his own identity and lived experience, or to share and risk all credibility. It is not possible for anyone outside of the dominant group to show up as fully themselves.

*B*lack scientists are black people. While this may seem obvious, racism is often verbally disregarded (or worse, silently affirmed by social norms) in STEM spaces which causes real harm. Despite the fact that many white scientists praise their field for being objective by nature, there is an abundance of examples of racism within STEM settings. First, though, we must consider the racial trauma that black scientists carry with them to work because they live in an oppressive society; the reality of racism in the United States cannot simply be left at the door of the laboratory.

In June 2019, the *Journal of Adolescent Health* published a study which found higher rates of depressive thoughts and post-traumatic stress disorder in black and Latinx adolescents who had been exposed to viral videos of violence against people of color (Tynes et al., 2019). In 2016, Dr. Monnica Williams for *PBS News Hour* explained how these videos affect her as an adult, and also how they seem *not* to affect her white counterparts (Williams, 2016). Black scientists are coming to work with greater stress due to their experiences with racism, which is often made worse by the ignorance, obliviousness, and insensitivity of their white colleagues (*Race and Epistemologies of Ignorance*, 2007). Even worse, many black scientists face blatant discrimination and microaggressions regularly in their workplaces.

In the documentary *Picture a Scientist*, Chemist Raychelle Burks described [an example](#) of what microaggressions look like in a science setting (Cheney & Shattuck, 2020). A white colleague once mistook her for a janitor while sitting at her desk, surrounded by academic papers. Based on the perceptions of this colleague, it is clear that Burks did not fit into their idea of what a chemist looks like. It was a way of saying that Burks did not belong. Situations like this can escalate into life-threatening ones if the police are called.

In Brown's *Physics Today* commentary, he described an interaction with members of the Chicago Police Department, during his participation in a 10-week research internship at the University of Chicago (2020):

> *That night in the park overlooking the lake, I was aggressively approached by two police officers. One of the officers had his hand on his pistol and was yelling profanities at me; the other was muttering inaudibly. 'This is how I die,' I thought to myself. One officer* *slammed my head onto the hood of the police cruiser and painfully handcuffed me. Heart pounding and ears ringing, I found myself staring deeply into the eyes of an onlooking white couple who were sitting on a park bench, maybe 20 feet away. Apparently, a curfew was in effect—I was unaware because there were no posted signs—and the white couple and I were both in violation of it. Yet as I was being dehumanized and humiliated, one of the officers calmly walked to the white couple and politely asked them to leave. Before getting released, I spent 30 minutes in the small space in the back of the police cruiser, handcuffs cutting into my wrists, with no explanation of why I was being detained.*

It's not hard to imagine how this impacted his nightly run to the lakefront and back, and more broadly his research experience that summer. Though this incident technically happened off campus, there is little doubt that he'd associate it with the university. Imagine how he felt entering the lab the next day. While he "fabricate[d] thin films of quantum dots" he also tried to process, "This is how I die." Aside from the double-mindedness of his experience alongside the couple who witnessed it, he must have wondered what his white peers would have thought, and wrestled with whether telling them what happened would increase or diminish the alienation he already felt. In a setting where people broadly describe themselves as having a "culture of no culture," it is incredibly difficult for people of color to feel empowered or safe to speak about their experiences with the police, racism, and microaggressions within the setting: whether it is a classroom, laboratory or other professional space, common area, student union, athletic field, recreational space, or in the community surrounding the institution.

Siyonbola and Brown both remained in police custody for a brief time and had no charges filed against them. However, even "brief" minutes in police custody added stress and trauma to their journeys that long outlasted the duration of the encounter itself. That Ms. Siyonbola and Dr. Brown went on to pursue doctoral studies at prestigious institutions, University of Cambridge (UK) and Yale University, respectively, evinces their determination despite these encounters with policing (Brown II, 2020; *Yale Student Guilty of "Sleeping While Black" Graduates and Head to Another Prestigious University*, 2019). Data on police use of force, including killings and nonlethal injuries are extremely difficult to get, so determining the scope of direct impact on black scientists is hard to discern (Jackman, 2021). Harder still is getting data on how many police encounters happen in STEM spaces. Yet it's easy to see the connection. Whether faculty or graduate students, navigating the world while black, impacts black scientists in ways it does not for their white peers.

G<small>ATEKEEPING</small>
Physicist Charles Brown has a list of impressive achievements that date back to his time as an undergraduate student at the University of Minnesota. However, none of these achievements are visible just by looking at him, unlike the color of his skin. In graduate school at Yale University, Brown was one of only 25 black men enrolled and constantly made to feel like an outsider. He recalls being denied entry into certain buildings, while white students entered without showing ID, and being asked if he was "affiliated with Yale" while in student-only spaces (Brown II, 2020).

There is a perception that people of color, particularly black people, do not belong in STEM—this perception is made clear by Burks' story of being mistaken for a janitor, or how Charles Brown was repeatedly asked to show his student ID. These incidents cut into time that these scientists could be spending on grants, research, networking, and helping students (Cheney & Shattuck, 2020). Burks also describes how she has to spend time responding to inappropriate emails, and before she can respond how she must spend time thinking about the right approach to take, because expressing her anger is not an option for her as a black woman in STEM. Each incident may only require up to 20 minutes of her time, but consider how much that adds up to be over the years. While Burks doesn't explicitly say this, dealing with these interactions may also require time spent cooling down, decompressing, and refocusing on the work that she and other people of color were actually hired to do.

The time that a person of color in STEM uses to deal with oppression is time that white counterparts use to do the things mentioned above, that all academics want to devote time to. This difference makes white scientists look more productive, motivated, or dedicated. This perceptual difference also holds up the flawed but pervasive belief that people of color are lazy, and therefore lacking the work ethic required to be good scientists (Kachchaf et al., 2015; Turner, 2002). This belief is definitional of *the lie* (Glaude Jr., 2020). It is pervasive, despite the fact that this nation was literally built on the deaths and lands of Native people; and by enslaved people, largely of African descent.

Yet the idea that white men are better scientists than others is widely held in American society. This is the lie in scientific sheep's clothing. Our societal buy-in to this way of thinking, leaves scientists of color at a disadvantage because they are thought of automatically, unconsciously, as being less hard-working. Theoretical physicist Clifford V. Johnson, for example, said in an October 2020 interview with *Physics Today* that he has "had to run faster and jump higher and achieve more just to be treated the same as everyone else." He feels that his blackness is the unspoken reason for being excluded from opportunities (Feder, 2020):

> *I've sat on review panels for funding agencies, and I've seen how those panels work when I am in the room, with strongly stated and often unsubstantiated opinions swaying decisions. It does not fill me with confidence about what happens when I'm not in the room and I am the one being discussed.*

Johnson described himself as "spectacularly underfunded" at every stage of his career; he has never been able to support a postdoc (Feder, 2020). This, coupled with the decision-making he has witnessed on review panels, absolutely gives reason for concern and lack of confidence. Meritocracy has thus become a blatant lie that holds white supremacy firmly in place.

Theoretical physicist Dr. Chanda Prescod-Weinstein challenges the lie in her 2019 article "Making Black Women Scientists under White Empiricism: The Racialization of Epistemology in Physics." In it, Prescod-Weinstein introduces the concept of white empiricism, defined as "the phenomenon through which only white people (particularly white men) are read as having a fundamental capacity for objectivity and Black people (particularly Black women) are produced as an ontological other" (Prescod-Weinstein, 2020, p. 421). This means that black women are routinely dismissed when they try to discuss *their own experiences* of racism, both in a physics setting and in society at large.

Considering that empiricism is the theory that all knowledge comes from the senses, white empiricism is therefore an "antiempirical disposal of data" (Prescod-Weinstein, 2020, p. 423). Arguments in the diversity, equity, and inclusion (DEI) space often ask the audience to consider how much the sciences are missing due to this disposal and dismissal of black women and groups who are othered (Gibbs Jr., 2014; Graves et al., 2022; Hong & Page, 2004;

Jimenez et al., 2019; Ouimet, 2015). While it is true that the physics community as a whole loses when it routinely bars certain demographics from entering and contributing, what matters here is the black women *deserve* a seat at the table for no reason other than they have an interest they want to pursue. BIPOC people deserve to be welcomed to and respected in the physics space, and any space they may choose to inhabit.

However, welcoming and respecting must be the first steps taken, not the only steps. In the same aforementioned article, Dr. Prescod-Weinstein also addresses prestige asymmetry, a concept introduced by Joseph Martin in 2017. Martin notes how results from high energy physics are thought of as "intellectual achievements" meanwhile condensed matter is much less valued (Martin, 2017; Prescod-Weinstein, 2020, p. 423). He attributes this difference in assigned value to each field as the reason why high energy physicists receive greater resources, support, and recognition for their work when compared to condensed matter physicists. Sharon Traweek's work predates Martin's concept so her language is different, but Prescod-Weinstein argues that her understanding of how prestige functions in physics is more accurate. Traweek argues that "prestige in physics is determined by finely tuned social dynamics through which scientists' identities are made" (Prescod-Weinstein, 2020, p. 423; Traweek, 1988). In the so-called "culture of no culture" which physics claims to be, the demographics of the prestigious high-energy discipline is so homogenous, white and male, which by default creates a culture of sameness and even exclusion of those who are different.

It is worth noting that, of the less-than two hundred black American women to complete a PhD in Physics, the majority of these degrees were in areas that are not on the desirable side of prestige asymmetry (Prescod-Weinstein, 2020, p. 425; Valentine-Miller, 2021). This should be alerting us, as a physics community and a nation, that black women are being kept out of these prestigious fields not for lack of interest or competence but for lack of acceptance and civility. Even with these inconsistencies brought to light, the unbalanced support continues; even with the spotlight on racism and Black Lives Matter for much of 2020, the denial of racism in many spaces continues.

The work ahead is to confront the lie. We fail to make STEM inclusive because we do not see people who are not white men as equally worthy of our time and effort. Research shows that STEM faculty are far more aware of equity issues specific to white socioeconomically privileged women than of other groups, leaving them woefully deficient to address the range of intersectional gender inequity in their departments (Dancy & Hodari, 2021; Mervis, 2022a, 2022b). Thus, we must reorient to fixing ourselves, rather than the people we have historically excluded.

As was discussed in the *Informal Socialization* paper, as mentors, we must have some skin in the game (Hodari et al., 2022a). We must risk something real in order to be effective. Evidence has shown that we are far more likely to do this for people who are most likely us (by race, gender identity, class, orientation, or any of our identities), or for people who are already successful, or both. This is gatekeeping, policing the borders of our discipline. And when the discipline is almost entirely white men, this ensures that it will remain this way.

From this perspective, it is easy to see that gatekeeping in all its forms follows the contours and functions of policing: it uses power (violence) to hold the lie firmly in place. These many forms of gatekeeping act to undermine all but the most privileged, by:

- Continuously signal to women and BIPOC people that they do NOT belong, and waste their time and energy responding to those signals and other nonsense related to our belief in the lie;

- Inaccurately assess the accomplishments of women and BIPOC people as less than those of white men, and compound this by constraining their resources;
- Ignore testimonial injustice and prestige asymmetry, and feign innocence as women and BIPOC people are shut out of the premium reward structures;
- Display unwillingness to acknowledge white male privilege, and thereby use epistemic injustice to undermine any attempts to create more just disciplinary and institutional cultures; and
- Working hard to remain unseeing and unknowing about oppression, and thereby render yourself helpless, leaving all the work of dismantling oppression to others.

In the end, these behaviors delineate the firm boundary between those we deem worth the risks to our reputation and the effort of our investments, and those we simply do not. These behaviors wordlessly shout, "I'm not going out of my way to help people who don't belong here anyway!" even when our sincere intention says otherwise. But the impacts bely our intention, and our actions send a clear message to those who bear the brunt of them, as well as those few who explicitly intend harm. The latter can rest comfortably in the knowledge that we will remain stuck wrestling with our good intentions, unable and unwilling to make any real or lasting change.

*How to Help, Not Hinder*

So what can be done against such entrenched and systematic obstacles, and our unwillingness to oppose them? Some physics faculty have taken solid steps to ensure the success of all their students, as research on departments in which women of color thrive has displayed (Johnson, 2020; Johnson et al., 2017) In one study of white men physicists' knowledge and beliefs about gender and race, Cubit, a white man, and a full professor of physics described what he and his colleagues have done to address inequities in their department (Dancy & Hodari, 2021):

> *It's really all levels. Obviously you need some top down, looking at hiring practices, but we also just need to be- some of the things that we're trying to do, starting with addressing issues of sexism and racism and inclusion and diversity in our classes with students even in the first semester are just getting them used to thinking about it. So yeah, a bottom up approach too. I think both are certainly needed.*
>
> *One thing that, again, this hasn't been my idea, but something that I think some of the folks that I work with here have been particularly good about is making sure that we are addressing the* [white] *students and making sure to help combat exclusionary behaviors on their part. That's in some ways the most important thing, rather than- it's a complimentary part. I don't know if it's the most important part, but it's something that's often missing.*

Cubit's solutions may seem simple, but their impact has been profound. Separate interviews of his students reveal that he and his colleagues not only "help combat exclusionary behaviors" of white students' but the faculty interrupt sexist speech and behavior as it's happening (Hodari, 2018; Johnson, 2018). Students described their white man professor stopping his lecture to address a sexism comment, noting that even if he "was nerdy and uncomfortable" while doing this, the fact that he was willing to do it anyway made all the difference, and showed the women students that they could trust him to have their backs (Hodari, 2018; Johnson, 2019). These behaviors on the part of faculty open doors for their students rather than close them. They offer students emotional and psychological space, and transform the department into a counterspace, rather than a space students need to seek counterspaces from (Ong et al., 2018).

## REAL DIVERSITY, EQUITY, AND INCLUSION REQUIRE WHITE DISCOMFORT

White people are so uncomfortable with anti-racist work because of the deeply-held belief that only "bad people" can be racist and enact racial harm (DiAngelo, 2021, p. 13). This means that, even at superficial DEI talks and workshops, white people feel great discomfort and perhaps as if their character is under attack. This discomfort is often vocalized in a disruptive and unproductive manner, leaving the attendees of color feeling silenced and discouraged (Oluo, 2019). White people who volunteer and pay to attend such DEI events, then, seem to be doing so to feel better about themselves rather than to actually reduce the racial harm that they regularly perpetuate. With this in mind, it is inevitable that true, sweeping, systemic change can only be brought about at the expense of white feelings and white egos.

One of the greater challenges presented here is that white people tend to conflate their feelings and egos with their safety, and that white feelings and egos are particularly sensitive when racism is being discussed. Claims of white people, particularly white women, *feeling* unsafe has historically led to black people, including children like Emmett Till, Trayvon Martin, and Tamir Rice, to actually *being* unsafe (an understatement considering that each of these boys was murdered). Complaints about lack of safety do not always have lethal consequences; in the physics setting these consequences more likely result in shutting down efforts to create an inclusive workspace or university department. For example, Brian Nord experienced accusations of creating an uncivil work environment by a junior white colleague when he disagreed with them (calmly) in electronic communication about procedural matters and basic concepts in racism and EDI. Ironically, the accusations continued despite Nord's deliberate effort maintain a reasonable and measured tone. He highlighted how the dynamic of white *perceptions* of unsafety leading to actual unsafe conditions for a person of color (Nord) was playing out in their present interaction. This is often the overarching cause of people of color leaving STEM - whether for another field or leaving academia entirely.

While physicists may not be losing their lives as a result of anti-blackness and racism in the field, their struggle is a deep loss. It is a loss of identity as a fuller contributor to their chosen profession. It is the loss of return on a difficult and carefully considered investment of time and effort. There is a gap between the two realities, as experienced by a black physicist, and this may cause them material loss and psychological trauma.

Unfortunately, bad behavior of white people in DEI settings often does just this. Executive coach Yael Sivi explains in Forbes that psychological safety is "the sense on a team that one won't be embarrassed, shamed, or punished for their opinions and ideas" (Sivi, 2020). When white people act out in DEI settings, their colleagues of color feel psychologically unsafe as a direct result. While white people use the words "I feel unsafe" with nauseating frequency (DiAngelo, 2021, p. 120), the lack of psychological safety among people of color is supported by evidence. Oluo described multiple occasions of black women explicitly telling her that their workplace was not the right setting to pursue anti-racism. One woman did not even feel safe to say this out loud; instead, she wrote Oluo a note concluding that she "will heal at home in silence" (Oluo, 2019). People of color are at risk of harassment, loss of opportunities to advance their career, and social ostracization if they speak out about racism in the workplace; only white people have the luxury of being honest without repercussions in these situations despite the harm they cause. If white people would instead learn the truth about racism in this country, and be willing to speak about it, some real change may be possible.

While plenty of evidence exists to tell us that physics is not a welcoming environment for people of color, especially black women, the racial and gender disparity is quite alarming. Increasing the number of people of color in physics will not simply give these groups strength in numbers to solve the problem of discrimination. Rather, this issue gives white physicists an opportunity to be uncomfortable, and reflect on why it is that there are so few people of color at every level (especially underrepresented minorities such as black, Latinx, and Indigenous people). Dr. Charles Brown makes this challenge to the physics community at large:

> *Physics departments should not drive out Black students only to later say in bewilderment that "we just can't find the talent" and absolve themselves of responsibility for Black underrepresentation at each higher academic stage. Many departments require a restructuring to ensure that true value is placed on having Black students, staff, and faculty among their ranks. Such a restructuring should involve establishing policies and norms that are inclusive of Black physicists, enhance their sense of belonging, and place value on their academic contributions; importantly, it must also include building a diverse department by recruiting and developing more Black graduate students, postdocs, and professors* [(Chambers, 2017; Cochran, 2017)]. *Recruitment and development in an inclusive environment must be a package deal.*

From his perspective, it is up to the decision-makers, the admissions officers, department chairs, etc. to be willing to feel discomfort and look inward to admit that their decisions are skewed by racism and prejudice rather than blame the homogeneity of physics on a shortage of talented students of color.

Leaders need to ask themselves and the organizations they lead some hard questions. Why am I comfortable in a racist society? Why am I comfortable with the parallel oppressions in our organization? How might we distinguish between our discomfort and the actual lack of safety targets of oppression suffer? Staying alive is not the same thing as needing/wanting comfort. Whose safety are we centering? Whose security do our policies and practices secure? Whose sanctity is evidenced here? How do our behaviors, policies and praxis support the lie?

Wrestling with discomfort in this work requires ongoing effort. As this exchange between bell hooks and Laverne Cox exemplifies, the struggle to clarify the distinction between comfort and safety can be difficult to pin down:

| Laverne Cox: | *I think about the world that we live in that is so polarized. A world that we live in that we don't know how to have conversations without yelling at each other, when we have differences. And I think about your work, and how you talk so beautifully about creating safe spaces in the classroom, specifically in **Teaching to Transgress**, where folks feel safe to have conversations* [(hooks, 1994)]. *And I think a lot of that is about love.* |
|---|---|
| bell hooks: | *But I have to interrupt you, because I am actually am critical of the notion of safety in my work. And what I want is for people to feel comfortable in the circumstance of risk. Because I think that if we wait for safety, the bell hooks that wasn't sure that she could* |

> *get on the stage with Janet Mock would never have gotten on that stage. The bell hooks that was afraid of, "What if I used the wrong words?" "What if I say the wrong thing?" I would've stopped myself. So that, to me, I'm very interested in what does it mean for us to cultivate together? Community that allows for risk, the risk of knowing someone outside your own boundaries, the risk that is love. There is no love that does not involve risk. I'm a little weary, because white people love to evoke the safe spaces. [laughter and applause] And so I have a tendency to be critical of that. But I do believe that learning takes place in the harmonious space, the space that you and I are embodying tonight. For I'm not agreeing with you—*

Laverne Cox: *I like that "harmonious" space. I love that. But I think that for me it is that so many of the risks that I have taken in my life, I've had to have some kind of support …*

bell hooks: *Alright!*

Laverne Cox: *… behind me to let me know that I would be OK, and that if this didn't quite work out that they would still love me. That is really what I'm talking about. How we have—when we feel, like, that someone's not going to turn their backs on us or demonize us when we go and take that risk, and maybe use the wrong pronoun or say the wrong thing. I'm terrified of saying the wrong thing up here on stage with bell hooks. You know? But here we are. Here we are.*

bell hooks: *But I think that what allows us, then, to be in that space is loving kindness. It is what love is, that allows us to be in that space where we don't agree, where we may see things totally different. … So I think that one of the things that is the opposite of the safe space is to cultivate in one another the courage, the courage to be self-actualized. The realization that you have to choose to be who you are.*

While hooks and Cox have shared beliefs around love, as articulated in hooks' seminal work, they still wrestle to find a shared understanding of how to create a space in which to interrogate areas of disagreement. In the end, they leave safety (while acknowledging fear) for courage. They model for us how to move forward in this work and not let fear stop us. They push towards each other, "I love that." "Alright!" while acknowledging the fear. They thereby encourage—put courage in—each other so that they can subsequently talk about areas of disagreement, such as discord between feminist claims and personal freedoms.

Despite the friendliness of their words, this is not a small discordance. They articulate fault lines in perspective playing out across the nation in legal, personal and societal battles (Barnes, 2010; Kerr, 2013; Nathoo, 2020; West & Zimmerman, 1987). Yet they modelled courage in the face of discord, and they did so publicly, displayed on a youtube video with 148,000+ views.

If these women, both economically privileged yet each embodying multiple disadvantaged identities in an oppressive societal context, can stand up in themselves and engage, so can we. Those most privileged in societal and disciplinary contexts—white men leaders in physics—can use this example to wrestle with themselves and their peers over how to make our disciplinary spaces better. Specifically, how do we accept discomfort, leave behind the quest for safety, and find paths to courage despite continuing fear. Can we apply hooks' idea "to cultivate together … [c]ommunity that allows for risk"? Can physicists with the most power and privilege—white men—choose to risk getting it wrong, but asking the tough questions anyway, despite discomfort and fear? How do we set up structures and practices so that we no longer feel comfortable in a racist society, but instead we feel the motivating discomfort in the presence of oppression? And finally, how do we create these structures and practices so that they propel us to act.

Situating his experience in the well-documented reality of many black physicists, Charles Brown tells us (2020):

> *Black physicists are forced to deal with a toxic combination of persistent scrutiny and suspicion, fear of retribution for voicing our concerns, denial that bias and discrimination exist, and isolation—all while working to broaden and deepen human knowledge. And we are expected to do so with grace. That unsupportive environment emerges both systemically and interpersonally, and it leads to the disillusionment of Black physicists at all levels.*

These experiences sit alongside descriptions by white men physicists about their own experiences with race and gender in physics. In a recent interview study, David, a white man, a full professor of physics, and chair of his department, described his learning about racism and sexism taking place on his campus (Dancy & Hodari, 2021; Mervis, 2022b):

> *We had here at my university about four or five years ago a town hall meeting where a bunch of African American students got up and talked about their experience at my university, and it was not what I expected. It was very strange to hear that. It came to me as a jolt of lightning that actually caused me to start getting much more involved in these kinds of efforts.*
>
> *I was the typical physics professor, I couldn't care less about anything but my research and physics and stuff like that. I thought I was above the fray and I don't have to deal with these things. I realized that I cannot live like that anymore. I should first of all learn about these things a lot and be proactive. So I became a social activist. If you had told me ten years ago that I would become a social activist I would look at you with pity. So yes, this town hall for me was really a watershed moment.*
>
> *It was about four or five years ago. I don't remember exactly. It had nothing to do, by the way, with my department. The students who talked there were mainly in humanities or athletics. But they talked about systemic racism that took me by surprise, especially because our chancellor at the time was an African American woman. But putting*

> *one or two people here and there, scattering them around that have these kinds of credentials, does not really change the real culture of an institution. We unfortunately have an institution that is 90-something percent white in a state that is 90-something percent white. So we have a lot of work to do.*
>
> *One said, which struck me really bad, that he is from South Carolina. No, from Georgia, he was from Georgia. My family are actually from Georgia and I know that my grandmother was a total racist. He said that he's from Georgia and he has never felt so attacked because of his race here as he was in Georgia. Here it's worse than what he encountered in Georgia. That kind of blew my mind. Another person came and said that she's from Texas and the same kind of thing. Altogether horrifying stories about just ignoring, placating, trying to even more than what they already are marginalize them as a group.*

As many people experience (DiAngelo, 2018, 2021; DiAngelo & Menakem, 2020), David was surprised by the racist encounters students were having around his campus, though he is careful to say that these were not students in his department. He also used the well-worn proximity argument to further his surprise, although he immediately critiqued it as not an effective remedy to systemic oppression. Further he talked about a racist ancestor, presumably as a counterpoint to his newfound identity as a "social activist".

As was discussed in the *Power Dynamics* paper, David makes common moves DiAngelo describes: (institutional) credentialing in his mention of the black woman chancellor; and personal credentialing in his description of his grandmother's racism (Hodari et al., 2022b). He also distanced his sphere of influence (his department) by noting, "It had nothing to do, by the way, with my department. The students who talked there were mainly in humanities or athletics." This distancing continued later in the interview, when he explicitly named that he expected racist speech and behavior in places like South Carolina, despite the student narratives making it clear that they experienced more racism on his campus than in the southern states where they grew up.

David's story is not monolithically bad. He recognized that racism on his campus is systematic, and his desire to connect with his newfound activism seems genuine. But he also spent a considerable portion of the interview performing his own credentialing, by listing the DEI-related committees on which he sat without describing how he engaged with anti-oppression *actions* or *content*, a version of "out-woking" other whites DiAngelo describes. As previously discussed, this amounts to little more than garnering progressive social capital *with other white people*.

In contrast, Cubit's focus on explicitly teaching white students to "help combat [their own] exclusionary behaviors" aligns with others that put institutional and programmatic power behind faculty and administrative practice to make STEM culture truly inclusive, alongside high achievement (Maton et al., 2009; Sto Domingo et al., 2019). Such practices—and the research that documents them—hold promise that moving beyond the surface, and confronting the many ways we (oft unknowingly) perpetuate and support oppression may help us create true justice for all those who inhabit physics spaces.

ACKNOWLEDGEMENTS

The authors acknowledge the Heising-Simons Foundation Grant #2020-2374 which supported this work. We also thank all the non-author organizers for the 2021 A Rainbow of Dark Sectors conference: Regina Caputo, Djuna Croon, Nausheen Shah, and Tien-Tien Yu. We additionally thank Risa Wechsler for her organizational support.

REFERENCES

Barnes, R. D. (2010). *Outrageous Invasions: Celebrities' Private Lives, Media, and the Law*. Oxford University Press.

Brown, C., Gonzales, E., Esquivel, j., Phillips, C., Sanders, V. A., Walker, A. L., Simpson, F., Edwards, M., Ramson, B., McAuley, O., Williams, G., & Platt, J. (2022). *Black In Physics*. https://www.blackinphysics.org/
https://twitter.com/BlackinPhysics

Brown II, C. D. (2020). *Commentary: Disentangling anti-Blackness from physics*. https://physicstoday.scitation.org/do/10.1063/PT.6.3.20200720a/full/

Chambers, L. (2017). *A Different Kind of Dark Energy: Playing Race and Gender in Physics* Yale University].

Cheney, I., & Shattuck, S. (2020). *Picture a Scientist* M. Pottle, I. Cheney, S. Shattuck, O. Babalola, & I. Yadao; https://www.pictureascientist.com/

Cochran, G. L. (2017). Understanding and promoting diversity and inclusion in physics. *Sigma Pi Sigma Observer*, Winter. https://www.spsnational.org/the-sps-observer/winter/2017/understanding-and-promoting-diversity-and-inclusion-physics

Dancy, M. H., & Hodari, A. K. (2021). *Discourses white men use to maintain white and male supremacy in physics* National Association for Research in Science Teaching 94th Annual International Conference,

DiAngelo, R. (2018). *White Fragility: Why It's so Hard for White People to Talk about Racism*. Beacon Press.

DiAngelo, R. (2021). *Nice Racism: How Progressive White People Perpetuate Racial Harm*. Beacon Press.

DiAngelo, R., & Menakem, R. (2020). *Towards a Framework for Repair* [Interview]. The On Being Project. https://onbeing.org/programs/robin-diangelo-and-resmaa-menakem-towards-a-framework-for-repair/

Feder, T. (2020). Black voices in physics: Clifford Johnson. *Physics Today*. https://doi.org/10.1063/PT.6.4.20201023e

Gerken, T. (2018). *Police called after black Yale student fell asleep in common room*. https://www.bbc.com/news/world-us-canada-44068305

Gibbs Jr., K. (2014). Diversity in STEM: What It Is and Why It Matters. *Scientific American*.

Glaude Jr., E. S. (2020). *Begin Again: James Baldwin's America and Its Urgent Lessons for Our Own*. Crown-Random House.

Graves, J. L., Jr, Kearney, M., Barabino, G., & Malcom, S. (2022). Inequality in science and the case for a new agenda. *Proceedings of the National Academy of Sciences of the United States of America*, *119*(10). https://doi.org/https://doi.org/10.1073/pnas.2117831119

Hodari, A. (2018). *Co-creating inclusion in physics: Learning from what helps women of color thrive, Keynote* International Conference of Physics Education, Johannesburg, South Africa.

Hodari, A. K., Krammes, S. B., Prescod-Weinstein, C., Nord, B. D., Esquivel, J. N., & Assamagan, K. A. (2022a). Informal Socialization in Physics Training. *arXiv*, Article arXiv:2203.11518 (physics.soc-ph). https://urldefense.com/v3/__http://arxiv.org/abs/2203.11518__;!!P4SdNyxKAPE!XXRFNHuvJsnZ8iQwx7X_K7WW_Cb7iK7MTIIHHN-_fFuYCfwJyDg7m8xYCjq_p8o$

Hodari, A. K., Krammes, S. B., Prescod-Weinstein, C., Nord, B. D., Esquivel, J. N., & Assamagan, K. A. (2022b). Power Dynamics in Physics. *arXiv*, Article arXiv:2203.11513 (physics.soc-ph). https://urldefense.com/v3/__http://arxiv.org/abs/2203.11513__;!!P4SdNyxKAPE!QvHMgtZnM6pbL0RjRd2LTTqxa2sJqlCbKmYfN5DFxo65MYI82fw40F9q9B0HDbo$

Hong, L., & Page, S. E. (2004). Groups of diverse problem solvers can outperform groups of high-ability problem solvers. *Proc Natl Acad Sci U S A*, *101*(46), 16385-16389. https://doi.org/10.1073/pnas.0403723101

hooks, b. (1994). *Teaching to Transgress: Education as the Practice of Freedom*. Routledge.

Jackman, T. (2021). FBI may shut down police use-of-force database due to lack of police participation. *The Washington Post*.

Jimenez, M. F., Laverty, T. M., Bombaci, S. P., Wilkins, K., Bennett, D. E., & Pejchar, L. (2019). Underrepresented faculty play a disproportionate role in advancing diversity and inclusion. *Nature Ecology & evolution*, *3*, 1030-1033. https://doi.org/https://doi.org/10.1038/s41559-019-0911-5

Johnson, A. (2020). An intersectional physics identity framework for studying physics settings. In A. T. D. Allison J Gonsalves (Ed.), *Physics Education and Gender* (pp. 53-80). Springer Nature.

Johnson, A., Ong, M., Ko, L. T., Smith, J., & Hodari, A. (2017). Common challenges faced by women of color in physics, and actions faculty can take to minimize those challenges. *The Physics Teacher*, *55*.

Johnson, A. C. (2018). *"I can't really think of anything I don't like": Locating and learning from universities where women of color are thriving in physics*. St Mary's College of Maryland.

Johnson, A. C. (2019). *Creating environments where women of color can thrive in physics, Invited* American Physical Society March Meeting,

Kachchaf, R., Ko, L., Hodari, A., & Ong, M. (2015). Carer-Life balance for women of color: Experiences in science and engineering academia. *Journal of Diversity in Higher Education*, *8*(3), 175-191. https://doi.org/http;??dx.doi.org/10.1037/a0039068

Kerr, D. E. (2013). *Butch, femme, dyke, or lipstick, aren't all lesbians the same?: An exploration of labels and "looks" among lesbians in the US south* The University of Mississippi].

Laughland, O. (2019). Sandra Bland: Video released nearly four years after death shows her view of arrest. *The Guardian*. https://www.theguardian.com/us-news/2019/may/07/sandra-bland-video-footage-arrest-death-police-custody-latest-news

Martin, J. D. (2017). Prestige asymmetry in American physics: Aspirations, applications, and the purloined letter effect. *Science in Context*, *30*(4), 475-506.

Maton, K. I., Domingo, M. R. S., Stolle-McAllister, K. E., Zimmerman, J. L., & III, F. A. H. (2009). Enhancing the number of African Americans who pursue STEM PhDs: Meyerhoff Scholarship Program outcomes, processes, and individual predictions. *Journal of Women and Minorities in Science and Engineering*, *15*(1), 15-37.

Mervis, J. (2022a). *Black Colleges Can't Do It All*. https://www.science.org/content/article/u-s-black-colleges-train-outsize-share-physics-majors-they-can-t-do-it-all

Mervis, J. (2022b). The toll of white privilege: How dominant culture has discouraged diversity. *Science*, *375*(6584).

Nathoo, Z. (2020). *How can employees also be social-media activists?* The Life Project.

Oluo, I. (2019). *Confronting racism is not about the needs and feelings of white people*. https://www.theguardian.com/commentisfree/2019/mar/28/confronting-racism-is-not-about-the-needs-and-feelings-of-white-people

Ong, M., Smith, J. M., & Ko, L. T. (2018). Counterspaces for women of color in STEM higher education: Marginal and central spaces for persistence and success. *Journal of Research in Science Teaching*, *55*(2), 206-245. https://doi.org/https://doi.org/10.1002/tea.21417

Ouimet, M. (2015). 5 numbers that explain why STEM diversity matters to all of us. *Wired*.

Prescod-Weinstein, C. (2020). Making black women scientists under white empiricism: The racialization of epistemology in physics. *Signs: Hournal of Women in Culture and Society*, *45*(2).

*Race and Epistemologies of Ignorance*. (2007). (N. T. Shannon Sullivan, Ed.). State University of New York Press.

Rogo, P. (2020). *The Black Yale Student Who Was Racially Profiled For Napping Speaks, Says University 'Has Not Done Enough'*. https://www.essence.com/news/black-yale-student-lolade-siyonbola-napping-while-black/

Sivi, Y. C. (2020, Apr. 6, 2020). *Psychological Safety isn't the Same As Being Comfortable*. https://www.forbes.com/sites/forbescoachescouncil/2020/04/06/psychological-safety-isnt-the-same-as-being-comfortable/?sh=ab4f915773d2

Sto Domingo, M. R., Sharp, S., Freeman, A., Freeman, T., Harmon, K., Wiggs, M., Sathy, V., Panter, A. T., Oseguera, L., Sun, S., Williams, M. E., Templeton, J., Folt, C. L., Barron, E. J., III, F. A. H., Maton, K. I., Crimmons, M., Fisher, C. R., & Summers, M. F. (2019). Replicating Meyerhoff for inclusive excellence in STEM. *Science*, *364*(6438).

Traweek, S. (1988). *Beamtimes and Lifetimes: The World of High Energy Physicists*. Harvard University Press.

Turner, C. (2002). Women of color in academe: Living with multiple marginality. *The Journal of Higher Education*, *4*, 74-93. https://doi.org/http://dx.doi.org/10.1353/jhe.2002.0013

Tynes, B. M., Willis, H. A., Stewart, A. M., & Hamilton, M. W. (2019). Race-Related Traumatic Events Online and Mental Health Among Adolescents of Color. *Journal of Adolescent Health*, *65*(3), 371-377. https://doi.org/10.1016/j.jadohealth.2019.03.006

Valentine-Miller, J. (2021). *AAWIP: The Physicists*. https://aawip.com/aawip-members/

West, C., & Zimmerman, D. H. (1987). Doing Gender. *Gender & Society*, *1*(2), 125-151. https://doi.org/10.1177/0891243287001002002

Williams, M. (2016). White people don't understand the trauma of viral police-killing videos. *PBS News Hour*. https://www.pbs.org/newshour/nation/column-trauma-police-dont-post-videos

*Yale Student Guilty of "Sleeping While Black" Graduates and Head to Another Prestigious University*. (2019). https://blacknews.com/news/lolade-siyonbola-yale-student-guilty-sleeping-while-black-graduates-another-prestigious-university/